\documentclass[sigconf]{acmart}

\usepackage{todonotes}
\usepackage{xspace}
\usepackage[capitalise]{cleveref}
\usepackage{balance}
\usepackage{booktabs}
\usepackage{tabularx}
\usepackage{subcaption}
\usepackage{comment}

\AtBeginDocument{%
	}

\copyrightyear{2025}
\acmYear{2025}
\setcopyright{acmlicensed}\acmConference[ISSTA Companion '25]{34th ACM SIGSOFT International Symposium on Software Testing and Analysis}{June 25--28, 2025}{Trondheim, Norway}
\acmBooktitle{34th ACM SIGSOFT International Symposium on Software Testing and Analysis (ISSTA Companion '25), June 25--28, 2025, Trondheim, Norway}
\acmDOI{10.1145/3713081.3731730}
\acmISBN{979-8-4007-1474-0/2025/06}

\begin{document}
	
	\title[Teaching Software Testing and Debugging with the Serious Game Sojourner under Sabotage]{Teaching Software Testing and Debugging with the Serious Game Sojourner under Sabotage}
	
	\author{Philipp Straubinger}
	\affiliation{%
		\institution{University of Passau}
		\country{Germany}}
	
	\author{Tim Greller}
	\affiliation{%
		\institution{University of Passau}
		\country{Germany}}
	
	\author{Gordon Fraser}
	\affiliation{%
		\institution{University of Passau}
		\country{Germany}}
	
	\renewcommand{\shortauthors}{Straubinger et al.}
	\newcommand{\toolname}{\emph{Sojourner under Sabotage}\xspace}
	\newcommand{\passau}{University of Passau\xspace}
	
	\newcommand{\summary}[2]{%
		\vspace{-0.2cm}%
		\begin{center}%
			\colorbox{gray!20}{%
				\parbox{\linewidth}{%
					\textbf{\textsf{Summary (\textit{#1})}:}~%
					#2%
				}%
			}%
		\end{center}%
	}
	
	\begin{abstract}
		Software testing and debugging are often seen as tedious, making them challenging to teach effectively. We present \toolname, a browser-based serious game that enhances learning through interactive, narrative-driven challenges. Players act as spaceship crew members, using unit tests and debugging techniques to fix sabotaged components. \toolname provides hands-on experience with the real-world testing framework JUnit, improving student engagement, test coverage, and debugging skills.
	\end{abstract}

	
	
	\maketitle
	
	\section{Introduction}
	
	Software testing and debugging are critical skills in software engineering, yet they are often perceived as tedious and secondary to programming by students~\cite{park2025exploring,garousi2020software}. Traditional teaching methods struggle to engage learners, leading to a gap between theoretical knowledge and practical application~\cite{7814898,DBLP:journals/csedu/McCauleyFLMSTZ08,DBLP:conf/issre/StraubingerF23}. To address this challenge, gamification, and serious games have been explored as potential solutions to enhance motivation and learning outcomes in software testing education~\cite{DBLP:journals/ieee-rita/QuinteroA23,DBLP:journals/jss/BlancoTCCRT23,DBLP:conf/cist/YamoulOMR23,DBLP:journals/ce/ConnollyBMHB12}.
	
	In this paper, we therefore present \toolname, a browser-based serious game designed to teach software testing and debugging in an engaging, interactive manner. Players take on the role of a spaceship crew member who must identify and fix sabotaged components using unit tests and debugging techniques. Unlike conventional educational tools, \toolname integrates a compelling story with structured programming challenges, allowing students to gain hands-on experience with testing concepts and Java testing frameworks.
	
	The results of an initial study indicate that students find \toolname enjoyable and educationally valuable, with measurable improvements in test coverage, debugging performance, and motivation compared to traditional methods. These findings reinforce the potential of serious games to complement software engineering education and foster essential software testing skills.
	
	\section{Background}
	
	Software testing is often included in general programming courses rather than taught separately, although it has been argued to integrate it early in curricula~\cite{DBLP:journals/jss/GarousiRLA20}. However, students often perceive it as tedious and redundant, posing challenges for educators~\cite{DBLP:journals/jss/GarousiRLA20,DBLP:conf/issre/StraubingerF23}. Motivation is crucial, as professional testers thrive on curiosity and creativity~\cite{DBLP:conf/esem/SantosMCSCS17}. Similarly, debugging is essential but often receives little instructional focus~\cite{DBLP:journals/csedu/McCauleyFLMSTZ08,DBLP:conf/wipsce/MichaeliR19}. Effective debugging requires domain knowledge and experience, yet novices struggle due to misconceptions, leading to errors~\cite{DBLP:conf/ace/LiCDLT19,pea1986language,DBLP:journals/hhci/BonarS85}.
	
	Gamification can improve engagement in testing education~\cite{DBLP:conf/mindtrek/DeterdingDKN11,DBLP:books/sp/23/Cooper23}, but poor design may cause frustration~\cite{DBLP:conf/hefa/TodaVI17,DBLP:conf/gamification/KappenN13}. Serious games offer immersive, goal-driven experiences to enhance learning~\cite{DBLP:journals/ce/ConnollyBMHB12,DBLP:books/sp/23/Cooper23} thus addressing this concern. Still, they require significant resources and supplement rather than replace traditional education~\cite{DBLP:conf/cist/YamoulOMR23}.
	
	Several serious games focussing on software testing and debugging have been proposed. For example, the Testing Game~\cite{DBLP:conf/fie/ValleTBM17} teaches functional, structural, and defect-based testing but lacks hands-on test writing. In contrast, Code Defenders~\cite{DBLP:conf/icse/CleggRF17,DBLP:conf/sigcse/FraserGKR19} engages students in writing JUnit tests and creating mutations but has usability concerns. Code Critters~\cite{DBLP:conf/icst/StraubingerCF23,DBLP:conf/icst/StraubingerBF24}, aimed at younger learners, simplifies testing concepts with block-based programming. Unlike these, \toolname targets undergraduate students, offering practical experience with real-world testing frameworks.
	
	For debugging, Gidget~\cite{DBLP:conf/vl/Lee14} teaches programming by having players fix buggy code through structured tasks. RoboBUG~\cite{DBLP:conf/icer/MiljanovicB17} guides students in debugging C++ code using techniques like print statements and code tracing. While \toolname does not teach programming, it incorporates story-driven tasks and structured debugging challenges inspired by these games, though it currently lacks features like step-by-step execution.
	
	\section{Sojourner under Sabotage}
	
	\toolname is an educational game where players take on the role of spaceship crew members who are unexpectedly awakened due to a cryogenic pod failure. Assisted by a robot companion, they must secure the ship by detecting and repairing sabotaged systems through unit testing. When a component fails, an alarm sounds, signaling the need for debugging. The game consists of seven levels, each requiring players to restore a different subsystem. Through navigation, object interactions, and robot assistance, players are guided through progressively complex challenges.
	
	\subsection{Gameplay}
	
	\begin{figure}[t]
		\centering
		\includegraphics[width=\linewidth]{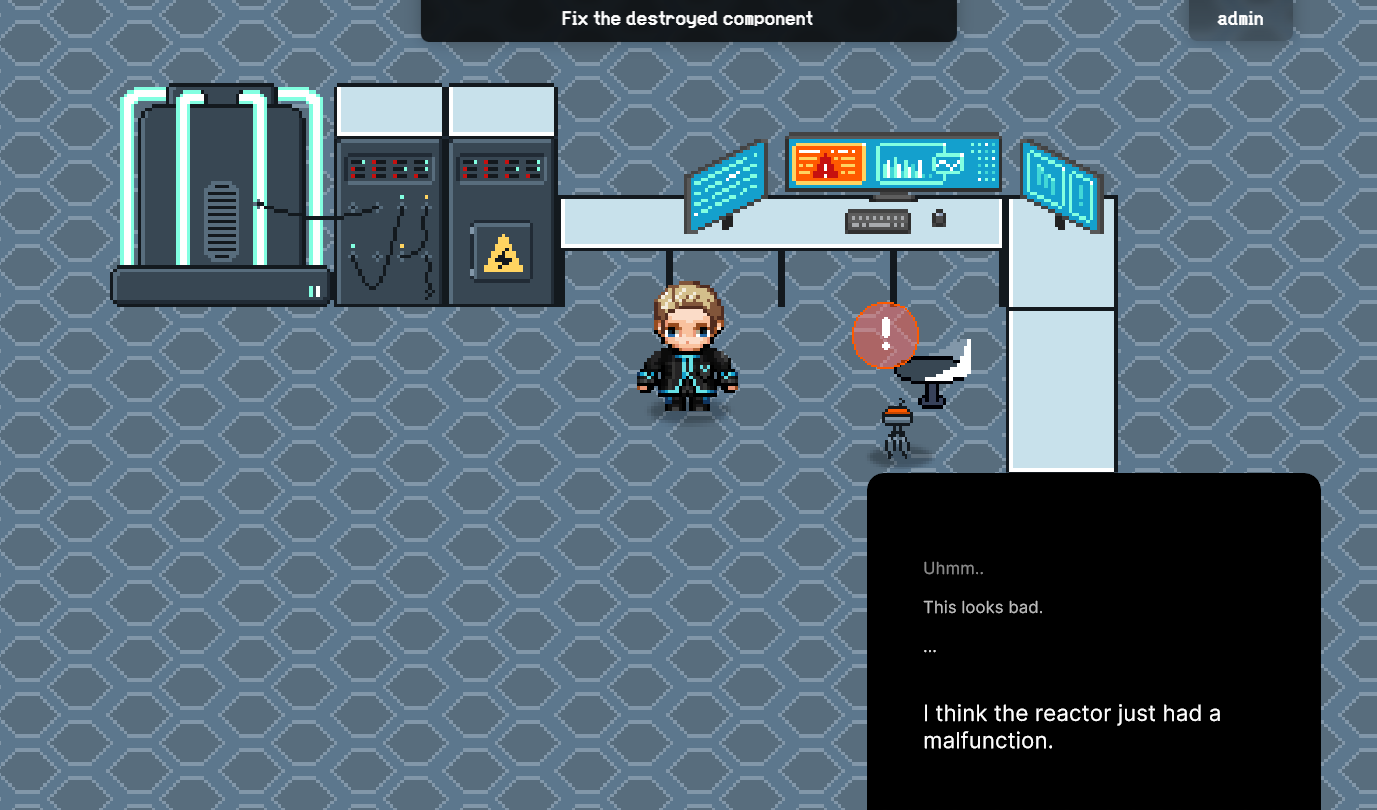}
		\caption{\toolname during an interaction between the player and the accompanying robot, which is reporting a sabotaged component}
		\label{fig:player}
	\end{figure}
	
	\begin{figure}[t]
		\centering
		\includegraphics[width=\linewidth]{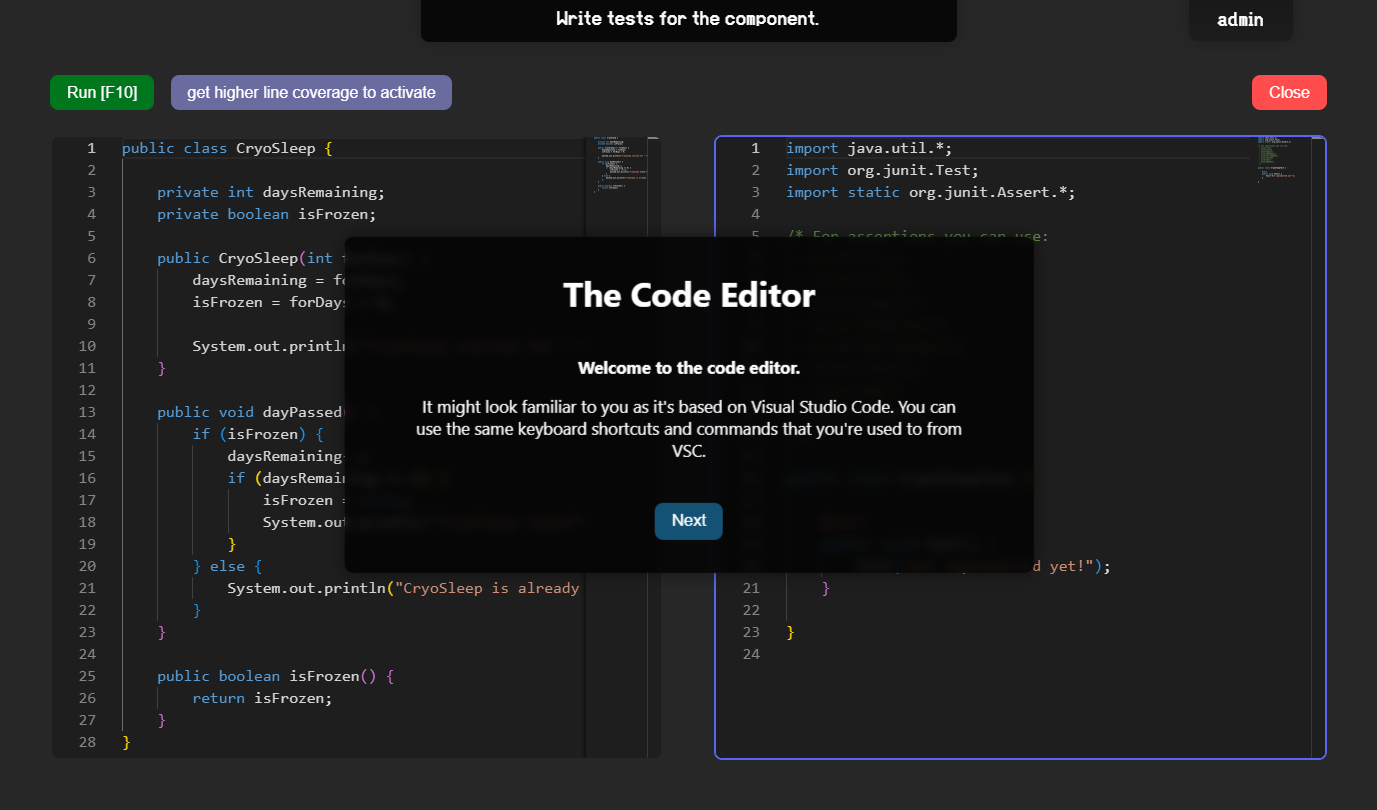}
		\caption{Code editor integrated into the game for writing tests (shown in picture), but also for debugging and fixing}
		\label{fig:editor}
	\end{figure}
	
	\begin{figure}[t]
		\centering
		\includegraphics[width=\linewidth]{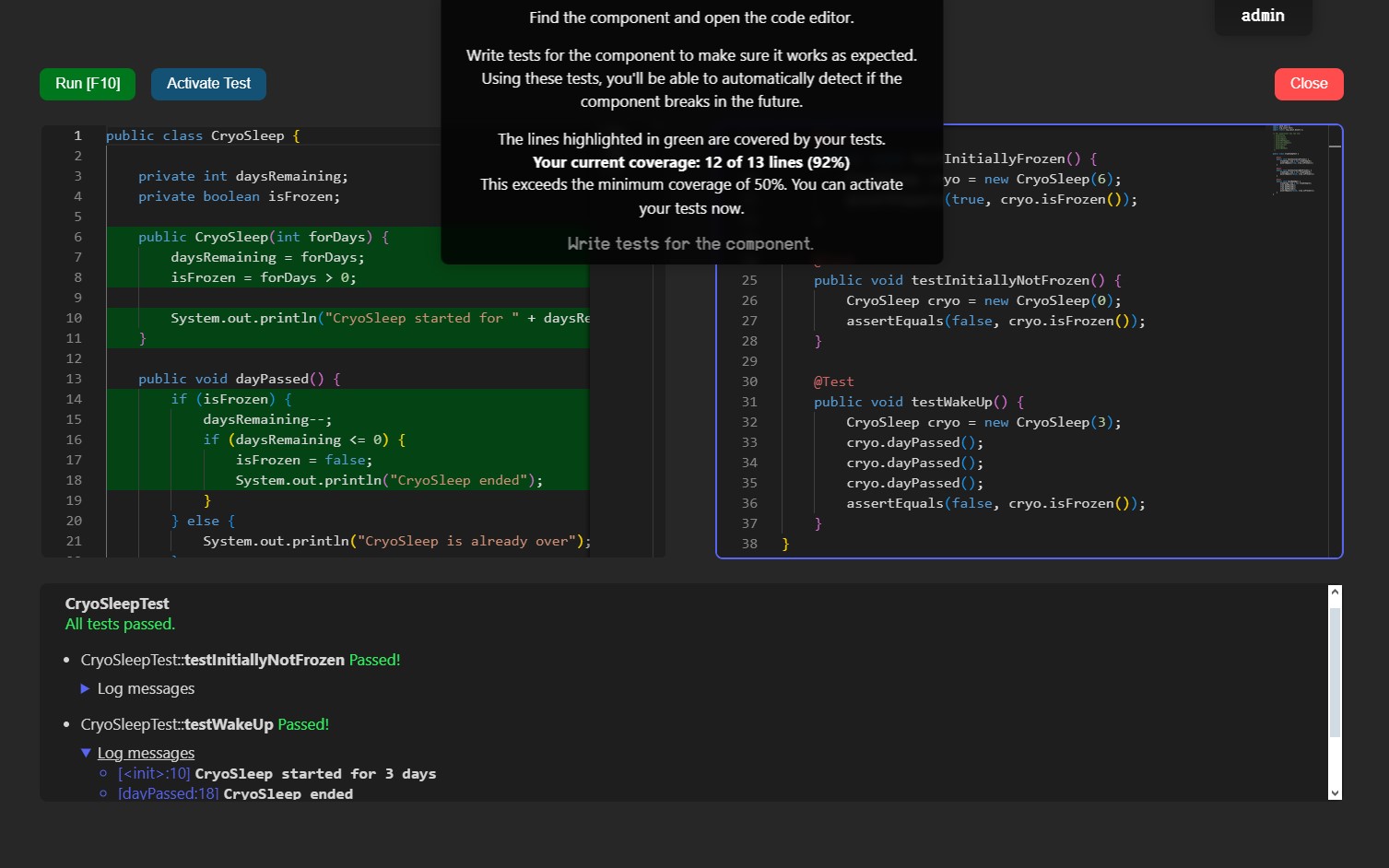}
		\caption{Code editor of \toolname after writing tests}
		\label{fig:finishedtests}
	\end{figure}
	
	\begin{figure}[t]
		\centering
		\includegraphics[width=\linewidth]{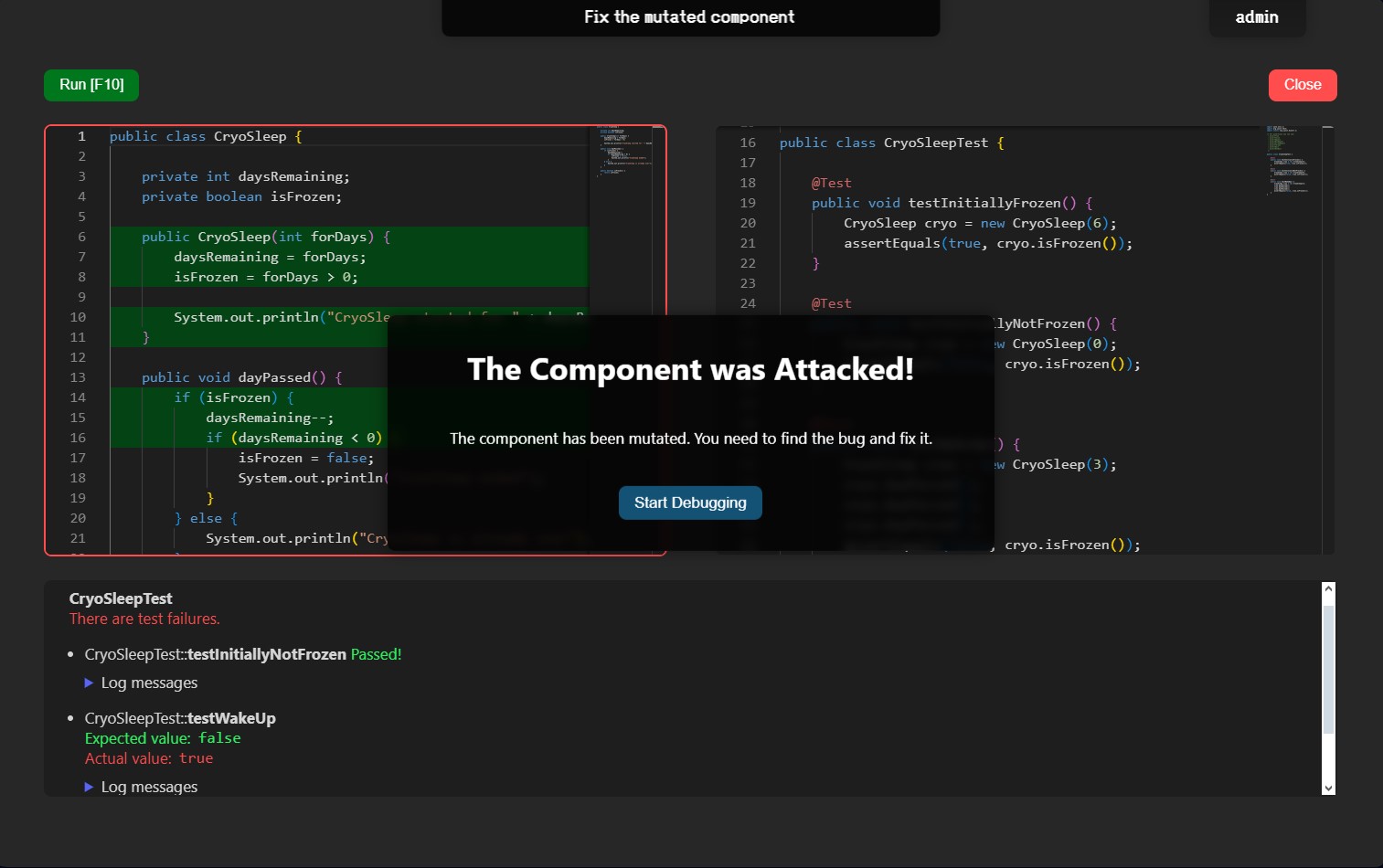}
		\caption{Code editor of \toolname while debugging}
		\label{fig:debug}
	\end{figure}
	
	The game begins with the player regaining consciousness in the spaceship environment shown in \cref{fig:player}, devoid of any memory of past events. The robot introduces itself and explains the situation, explaining the need to identify issues using unit testing. The first task involves locating the console, opening the onboard editor, and writing test cases (\cref{fig:editor}). Each level starts with the robot introducing a new room and its malfunctioning component. The player interacts with the component to access the code editor, where the testing process begins.
	
	Upon accessing the editor (\cref{fig:editor}), a tutorial explains its dual-panel layout: the left panel displays the component's source code, which must be tested and debugged if sabotage occurs, while the right panel is reserved for writing unit tests. Above the editor, buttons for executing tests, activating tests, and exiting the editor are available. Running tests provides immediate feedback in a console, indicating errors or successful executions. To advance, players must achieve at least 50\% line coverage before activating tests, allowing them to monitor for future sabotage automatically. Visual feedback reinforces the importance of test coverage, with successfully tested lines highlighted in green to indicate effective validation.
	
	Once tests are active, they continue running in the background until sabotage modifies the component's code. If an issue is detected, an alarm sounds, and the robot alerts the player (\cref{fig:player}). If sabotage goes undetected, the affected component sustains damage, prompting the robot to generate a guiding test case to assist in diagnosis. These supportive test cases help maintain engagement and prevent frustration, ensuring players remain focused on learning. At this stage, the player transitions from test writing to debugging, relying on error messages and failed test outputs to identify and fix issues.
	
	Returning to the console reveals sabotage-induced errors (\cref{fig:debug}). Debugging tools highlight problematic areas by providing logs, error messages, and code coverage visualization. Players analyze failures and correct the mutation, such as an altered comparison operator in the \texttt{dayPassed()} method. Additionally, players gain exposure to Java’s logging functionalities, strengthening their understanding of code analysis techniques.
	
	As players progress through the game, they unlock new rooms by completing levels and engaging in minigames. These circuit puzzles serve as cognitive breaks while maintaining thematic relevance. The overall seven levels form a continuous map, with each level introducing increasingly complex challenges that mirror real-world programming errors. The levels are structured around distinct programming concepts, including state management, floating-point arithmetic, iteration, data structures, and recursion.
	
	By blending problem-solving, testing, and debugging into a cohesive gameplay experience, \toolname fosters an interactive learning environment that strengthens players' confidence in writing and maintaining reliable software. Through immersive engagement with Java, unit testing, and debugging, players develop a deeper understanding of programming principles, reinforcing both theoretical knowledge and practical problem-solving skills.
	
	\subsection{Administrator Interface and Data Export}
	
	By default, an administrator account is created by naming an account \textit{admin}. Accessing \texttt{/admin} with this account enables Java Melody monitoring,\footnote{\url{https://github.com/javamelody/javamelody}} data extraction, and bulk user generation.
	
	The bulk user generation feature streamlines account setup for lectures. A print-friendly view is available for generated accounts.
	After a lecture, \toolname data can be downloaded in JSON format, including all recorded events and student actions. Fourteen event types are logged, each with metadata like an ID, username, and timestamp. Events differ in meaning and fields.
	
	When the game loads, the client signals readiness, and the server responds with the player's state: location (room number), active component, and room status (\textit{DOOR}, \textit{TALK}, \textit{TEST}, \textit{TESTS\_ACTIVE}, \textit{DESTROYED}, \textit{MUTATED}, and \textit{DEBUGGING}). Room phases occur sequentially, with updates during key transitions, such as moving from TALK to TEST, completing puzzles, or starting debugging.
	
	Every time a player modifies a test or a class under test, this action is logged along with the player's code.
	Several updates provide insights into the player's code based on test execution results. One update indicates a compilation error and includes the corresponding error message. If execution is successful, another update provides detailed results, including the class name, status (pass or fail), execution time, detailed line coverage information, and assertion errors for each test method if applicable. The same information is included in updates that occur when running tests against mutated code after sabotage.
	Another update occurs if a hidden test extends the player's test suite—triggered when an undetected bug is caught by at least one hidden test.
	
	
	\subsection{Implementation}
	
	The game is designed as a web-based application for cross-platform accessibility, leveraging Unity\footnote{\url{https://unity.com/}} and its WebGL\footnote{\url{https://www.khronos.org/webgl/}} build capabilities for smooth execution without installations. The backend, developed with Java Spring Boot,\footnote{\url{https://spring.io/projects/spring-boot}} provides an API, a WebSocket-based event system, and handles dynamic code execution using the Java Compiler API. The system is hosted on a Tomcat server\footnote{\url{https://tomcat.apache.org/}} with Hibernate\footnote{\url{https://hibernate.org/}} for database management. 
	
	User interactions occur via an HTML-based user interface (UI) and the seamlessly integrated Unity scenes. 
	The main part of the game uses a top-down 2D perspective, with the robot companion following the player via the A*~\cite{cui2011based} pathfinding and Reynolds’~\cite{reynolds1999steering} path-following algorithms.
	Some parts of the UI, like the dialogue boxes and the puzzle minigames, are implemented using Preact,\footnote{\url{https://preactjs.com/}} TypeScript, and Tailwind. They are then included in the Unity build via OneJS.\footnote{\url{https://onejs.com/}} Other parts, like the code editor and the pop-ups are developed in JavaScript and supplied as a webpack\footnote{\url{https://webpack.js.org/}} bundle to the website. The Unity game's C\# code communicates with these scripts by importing mediating JavaScript functions. This allows the usage of more powerful and prebuilt libraries like the Monaco Editor\footnote{\url{https://github.com/microsoft/monaco-editor}} which provides syntax highlighting, autocompletion, keyboard shortcuts and 
	customizations like the line coverage visualization. 
	
	A JSON API handles fetching, updating, and executing code.
	Events occurring in the game are transmitted using a bi-directional event system over WebSocket connections using the STOMP\footnote{\url{https://stomp.github.io}} protocol. 
	This is necessary additionally to the API, as some events, like the emergencies after a sabotage, are triggered by the server.
	All of these endpoints are secured by Spring Security in combination with JSON Web Tokens (JWT) to ensure controlled access.
	
	Gameplay relies on dynamic Java code execution, involving multiple steps: The test class and class under test (CUT) are fetched, compiled in-memory, instrumented, and loaded via a secure class loader. A JUnitCore\footnote{\url{https://junit.org/junit4/javadoc/latest/org/junit/runner/JUnitCore.html}} instance and execution thread are created with a timer to prevent infinite loops. The test class is executed, and the result, including coverage and logging data, is returned.
	
	\subsection{Deployment}
	
	The simplest way to deploy the game is by installing a WAR archive on a Tomcat server. The game has been successfully tested with Tomcat 10.1.23 and Java 17, but newer versions should work as well.
	To run the web application, a database is required. While various SQL databases are compatible, MySQL Community Server 8 is recommended. You can configure the database authentication details in the \texttt{application.properties} file, which is located in the \texttt{WEB-INF/classes} directory within the WAR file or the extracted archive.
	Finally, the application must be accessible from the desired domain. This can be done using a reverse proxy such as Nginx.\footnote{\url{https://nginx.org/}} A sample configuration file, typically found at \path{/etc/nginx/sites-enabled/tomcat.conf}, is included with the WAR archive.
	For secure connections, it is recommended to use a tool like Certbot\footnote{\url{https://certbot.eff.org/}} to configure an SSL certificate and enable HTTPS.
	
	\subsection{Extensibility}
	
	The game can be easily adapted to use different CUTs without requiring any modifications to the Unity export or the Java server code. However, it currently only supports Java as the programming language for components. The Unity source code can be used to create a new export, allowing changes to the rooms or even a complete redesign of the game's appearance.
	
	To streamline the development process, the game utilizes existing packages and assets. Two of these assets are licensed from the Unity Asset Store: OneJS by DragonGround LLC\footnote{\url{https://assetstore.unity.com/packages/tools/gui/onejs-221317}} and the RPG Map Editor by Creative Spore.\footnote{\url{https://assetstore.unity.com/packages/tools/game-toolkits/rpg-map-editor-25657}} Due to licensing restrictions, these assets are not included in the public repository. If a third party wishes to extend the game and build the Unity export themselves, they must obtain the appropriate licenses and place the packages in the \texttt{Assets} directory. However, the compiled export of the game, which includes these packages, is freely provided.
	
	The game's environment can be modified directly in Unity using the RPG Map Editor. The spaceship consists of a single large tilemap, making it intuitive to replace existing rooms or add new ones. It is also possible to change the entire setting of the game by using a different tileset with alternative textures.
	
	Beyond visual and environmental modifications, the room progression and the component's code can also be customized or expanded. The automatic data initialization service allows these modifications to be made directly in the resource files, ensuring that the database is updated accordingly. Adjustments to the order, number, and names of rooms, as well as the wait time between them, can be made by modifying the \texttt{game/game-progression.csv} file. The source code for a component is defined in the \path{cut/ComponentName.java} file, while mutated code and hidden tests follow the same structure and are located in the \texttt{mutants/} and \texttt{test/} folders.

	\section{Evaluation}
	
	To assess \toolname, we conducted controlled experiments with students from two different courses: a first-year software engineering (SE) course (45 participants) and a third-year software testing (ST) course (34 participants). Sessions were held in May and November 2024, where students played \toolname for 60 minutes, followed by a survey. Participants could seek external programming references but were restricted from using AIs. More details on this evaluation can be found in a research paper~\cite{Straubinger2025Sojourner} about the approach and the study.
	
	\subsection{Engagement and Performance}

	Participants primarily focused on writing tests and exploring the game, spending significantly less time debugging. The ST group, with greater programming experience, progressed further and performed better in debugging and testing, while less experienced SE participants required more time to advance. The time spent on testing was consistently higher than debugging, with ST students averaging 41 minutes and SE students 35 minutes. Both groups exceeded 50\% line coverage across levels, with ST students generally achieving better coverage and mutation scores. However, their higher number of tests also led to more test smells.
	
	In debugging, participants initially focused on understanding the code and executing tests before shifting to code modifications in later levels. The ST group demonstrated more caution, introducing fewer bugs, while SE participants were more prone to errors. The print debugging feature was rarely used, suggesting the need for more accessible debugging tools.
	
	\subsection{Student Perception}
	
	Feedback on \toolname was overwhelmingly positive. Over 80\% of participants enjoyed the game, praising its storyline, graphics, and mini-games. More than 90\% found test writing enjoyable, while over 75\% also appreciated debugging. The ST group generally felt more confident navigating the game, whereas the SE group found testing easier as they progressed. A majority of participants agreed that \toolname helped them practice valuable skills, reinforcing its effectiveness as a learning tool.
	
	\section{Conclusions}
	
	This tool paper demonstrates \toolname, a browser-based serious game designed to teach software testing and debugging interactively and engagingly. Our evaluation with 79 students demonstrated that the game effectively motivates learners of varying experience levels, with over 80\% of participants enjoying the gameplay and recognizing its educational value. The results showed that students with more advanced testing knowledge performed better in terms of line coverage and mutation scores but also exhibited more test smells, highlighting areas for improvement in test-writing practices. Additionally, debugging features were underutilized, suggesting the need for more accessible tools.
	
	Moving forward, we plan to enhance \toolname by refining debugging tools, including step-through execution and breakpoints, to reduce reliance on print debugging. We will also incorporate better guidance for quality test-writing practices and explore adaptive difficulty mechanisms to tailor challenges for different skill levels. Additionally, we aim to improve the editor experience with a language server and introduce tutorial levels or level-specific hints during testing and debugging.
	
	We also plan to expand gameplay with new level types, such as those requiring mocking or handling dependencies. Other planned improvements include indicators for better orientation on the spaceship, revisiting previous components later in the game, and displaying reference solutions at the end of every level. Finally, we are considering adding more minigames to enrich the overall experience.
	Lastly, future evaluations will expand to a broader audience, including students from diverse institutions and professional developers, to assess the game’s effectiveness in various contexts.
	
	\section*{Tool Availability}
	A short demo of \toolname can be found at:
	\begin{center}
		\url{https://youtu.be/KU8Tu5Aeo88}
	\end{center}
	The deployable version used in this paper can be found at:
	\begin{center}
		\url{https://doi.org/10.6084/m9.figshare.28550858}
	\end{center}
	\toolname can be played at:
	\begin{center}
		\url{https://sojourner-under-sabotage.se2.fim.uni-passau.de/}
	\end{center}
	
	\balance
	
	\bibliographystyle{ACM-Reference-Format}
	\bibliography{bib}
	
\end{document}